\documentclass[twocolumn, twocolappendix]{aastex631}
\pdfoutput=1
\usepackage{newtxtext,newtxmath}
\usepackage[T1]{fontenc}
\usepackage{ae,aecompl}
\usepackage{natbib,textcomp,placeins,epsfig,graphicx}
\usepackage{color,amsmath,upgreek,listings,epstopdf,placeins}
\usepackage{sidecap}
\usepackage{tikz}
\usepackage[flushleft]{threeparttable}

 \newcommand{\z}{$|z|$}

\shorttitle{Low-Al disk stars}

\shortauthors{D. K. Feuillet et al.}

\date{Accepted XXX. Received YYY; in original form ZZZ}

\begin{document}


\title{An old, metal-rich accreted stellar component in the Milky Way stellar disk}
 
\correspondingauthor{Diane K. Feuillet}
\email{diane.feuillet@astro.lu.se}

\author[0000-0002-3101-5921]{Diane K. Feuillet}
\affiliation{Lund Observatory, Department of Astronomy and Theoretical Physics, \\
Box 43, SE-221~00 Lund, Sweden}

\author[0000-0002-7539-1638]{Sofia Feltzing}
\affiliation{Lund Observatory, Department of Astronomy and Theoretical Physics, \\
Box 43, SE-221~00 Lund, Sweden}

\author[0000-0002-0918-346X]{Christian Sahlholdt}
\affiliation{Lund Observatory, Department of Astronomy and Theoretical Physics, \\
Box 43, SE-221~00 Lund, Sweden}

\author[0000-0003-3978-1409]{Thomas Bensby}
\affiliation{Lund Observatory, Department of Astronomy and Theoretical Physics, \\
Box 43, SE-221~00 Lund, Sweden}




\begin{abstract}

We study the possibility that the Milky Way’s cool stellar disc includes mergers with ancient stars. 
Galaxies are understood to form in a hierarchical manner, where smaller (proto-)galaxies merge into larger ones. Stars in galaxies, like the Milky Way, contain in their motions and elemental abundances tracers of past events and can be used to disentangle merger remnants from stars that formed in the main galaxy.
The merger history of the Milky Way is generally understood to be particularly easy to study in the stellar halo. The advent of the ESA astrometric satellite Gaia  has enabled the detection of completely new structures in the halo such as the \textit{Gaia}-Enceladus-Sausage. However, simulations also show that mergers may be important for the build-up of the cool stellar disks.
Combining elemental abundances for $\sim 100$ giant branch stars from APOGEE\,DR17 and astrometric data from \textit{Gaia} we use elemental abundance ratios to find an hitherto unknown, old stellar component in the cool stellar disk in the Milky Way. We further identify a small sample of RR\,Lyrae variables with disk kinematics that also show the same chemical signature as the accreted red giant stars in the disk. These stars allows us to date the stars in the accreted component. We find that they are exclusively old. 

\end{abstract}


\section{Introduction}
Galaxies are understood to form in a hierarchical manner, where
smaller (proto-)galaxies/haloes merge into larger ones \citep{1993MNRAS.262..627L,2017ARA&A..55...59N}. Each galaxy will have a unique star formation history and hence also a unique chemical evolution history. Different elements are released to the
inter-stellar medium by stars with different masses and therefore on
different timescales \citep{1996snih.book.....A,1997NuPhA.621..467N}.
This means that the elemental abundances are able to tell us about the
past history of the gas that the stars formed out of. { Low-mass}
stars retain, in their outer atmospheres, an imprint of the
composition of the gas from which they formed
\citep{1989AIPC..183..168L,2002ARA&A..40..487F}. By analysing the
spectra of such stars we can trace the chemical evolution of a stellar
population over time.  One important
aspect of star formation is that it proceeds faster in heavier systems
and slower in smaller systems \citep{matteucci1986}. This in turn means
that smaller galaxies that merged with the Milky Way have different
elemental abundance patterns compared to the stars that formed in situ
in the large galaxy. Several elements are affected but it has been
shown that the abundance of aluminum relative to iron is empirically 
lower in local dwarf galaxy or accreted stellar populations over a range of iron abundances
({ $-2.0 < [$Fe/H$] < 0.0$,} \citealp{2015MNRAS.453..758H,2019ApJ...872...58H, 2021ApJ...923..172H}). Thus, we infer that
stars in galaxies, like the Milky Way, contain in their motions and
elemental abundances tracers of past events and can be used to
disentangle merger remnants from stars that formed in the main galaxy.

The merger history of the Milky Way is generally understood to be
more visible in the stellar halo as the fraction of in situ stars is lower than in the disk
\citep{2020ARA&A..58..205H}. The advent of the ESA astrometric
satellite \textit{Gaia} \citep{2018A&A...616A...1G} has enabled astronomers to
find completely new structures in the halo, such as the
\textit{Gaia}-Enceladus-Sausage that is believed to be the dissolved stellar population of a massive accretion event
\citep{2018Natur.563...85H,2018MNRAS.478..611B,2019MNRAS.488.1235M,2020MNRAS.497..109F}.

\begin{table*}
\footnotesize
\begin{threeparttable}
\caption{APOGEE fields removed from consideration \label{tab:fields}}
\centering
\begin{tabular}{l l l l l l l l l}
\hline \hline 

\verb|47TUC| & \verb|ANDR*| & \verb|BOOTES1| & \verb|Berkeley*| & \verb|CARINA| & \verb|COL261| & \verb|CygnusX*| & \verb|DRACO| \\
\verb|FL_2020| & \verb|FORNAX| & \verb|GD1-*| & \verb|IC342_NGA| & \verb|IC348*| & \verb|INTCL_N*| & \verb|JHelum*| & \verb|LAMBDAORI-*| \\
\verb|LMC*| & \verb|M10| & \verb|M107| & \verb|M12-N| & \verb|M12-S| & \verb|M13| & \verb|M15| & \verb|M2| \\
\verb|M22| & \verb|M3| & \verb|M3-RV| & \verb|M33| & \verb|M35N2158| & \verb|M35N2158_btx| & \verb|M4| & \verb|M5| \\
\verb|M53| & \verb|M54SGRC*| & \verb|M55| & \verb|M5PAL5| & \verb|M67*| & \verb|M68| & \verb|M71*| & \verb|M79| \\
\verb|M92| & \verb|N1333*| & \verb|N1851| & \verb|N188*| & \verb|N2204| & \verb|N2243*| & \verb|N2264| & \verb|N2298| \\
\verb|N2420| & \verb|N2808| & \verb|N288| & \verb|N3201*| & \verb|N362| & \verb|N4147| & \verb|N5466| & \verb|N5634SGR2| \\
\verb|N5634SGR2-RV_btx| & \verb|N6229| & \verb|N6388| & \verb|N6397| & \verb|N6441| & \verb|N6752| & \verb|N6791| & \verb|N6819*| \\
\verb|N752_btx| & \verb|N7789| & \verb|NGC188_btx| & \verb|NGC2420_btx| & \verb|NGC2632_btx| & \verb|NGC6791*| & \verb|NGC7789*| & \verb|ORION*| \\
\verb|ORPHAN-*| & \verb|Omegacen*| & \verb|PAL*| & \verb|PLEIADES*| & \verb|SCULPTOR| & \verb|SEXTANS| & \verb|SGR*| & \verb|SMC*| \\
\verb|Sgr*| & \verb|TAUL*| & \verb|TRIAND-*| & \verb|TRUMP20| & \verb|Tombaugh2| & \verb|URMINOR| & \verb|moving_groups| & \verb|ruprecht147| \\
\verb|sgr_tidal*| & & & & & & & \\
\hline
\end{tabular}
\vspace{0cm}
\begin{tablenotes}
\item Note. --- Strings corresponding to the APOGEE~DR17 parameter {\tt FIELD}. The asterisk (*) is used as a wild card character, indicating multiple possible names.
\end{tablenotes}
\end{threeparttable}
\end{table*}

\begin{table*}
\footnotesize
\begin{threeparttable}
\caption{APOGEE programs removed from consideration \label{tab:progname}}
\centering
\begin{tabular}{l l l l l l l l l}
\hline \hline 
\verb|Drout_18b| & \verb|Fernandez_20a| & \verb|Geisler_18a| & \verb|beaton_18a| & \verb|cluster_gc| & \verb|cluster_gc1| & \verb|cluster_gc2| & \verb|cluster_oc| \\
\verb|clusters_gc2| & \verb|clusters_gc3| & \verb|geisler_18a| & \verb|geisler_19a| & \verb|geisler_19b| & \verb|geisler_20a| & \verb|halo2_stream| & \verb|halo_dsph| \\
\verb|halo_stream| & \verb|kollmeier_19b| & \verb|magclouds| & \verb|monachesi_19b| & \verb|sgr| & \verb|sgr_tidal| & \verb|stream_halo| & \verb|stutz_18a| \\
\verb|stutz_18b| & \verb|stutz_19a| & & & & & & \\
\hline
\end{tabular}
\vspace{0cm}
\begin{tablenotes}
\item Note. --- Strings corresponding to the APOGEE~DR17 parameter {\tt PROGRAMNAME}. We list only unique programs but some stars belong to multiple programs.
\end{tablenotes}
\end{threeparttable}
\end{table*}

However, simulations show that mergers may be important also for the
build-up of the cool stellar disks \citep{2008MNRAS.389.1041R, 
2017MNRAS.472.3722G}. So far, the empirical evidence for significant
merger remnants present in the Milky Way stellar disk is inconclusive
\citep{2014MNRAS.444..515R,2015MNRAS.450.2874R}. With the arrival of
the \textit{Gaia} parallaxes and proper motions for over a billion stars
\citep{2018A&A...616A...1G} in combination with radial velocity data
for bright stars from \textit{Gaia} and for fainter stars from ground-based
stellar spectroscopic surveys, it is now feasible
to study how stars move in the Milky Way. Such studies have successfully found accreted 
populations in the stellar halo, such as \textit{Gaia}-Enceladus-Sausage and many smaller
systems, as well as new streams in the
disk \citep{2020NatAs...4.1078N,2020A&A...639A..64R}. These
potential mergers present in the disk are either on non-circular
orbits, very young, or present only in the outer parts of the disk. To
find mature merger remnants in the stellar disk with disk-like,
circular orbits is a more subtle task where we need additional
information in the form of elemental abundances
\citep{2015MNRAS.450.2874R,2020NatAs...4.1078N}. Such data is provided
by large ground-based, high-resolution, spectroscopic surveys, such as APOGEE \citep{2017AJ....154...94M},
the \textit{Gaia}-ESO Survey \citep{2012Msngr.147...25G, 2013Msngr.154...47R}, and GALAH \citep{2015MNRAS.449.2604D}.

In this paper we make use of the astrometric data from the
\textit{Gaia} satellite in combination with data from APOGEE and the
literature to hunt for accreted populations in the stellar disk.  We supplement the APOGEE observations with a small sample of disk RR~Lyrae stars, { which are known to be old,} that have low measured aluminum abundances { and so may be associated with an accreted population, giving an estimate of age} \citep{2013RAA....13.1307L}.


\section{Data and methods}

\subsection{Stellar data from APOGEE DR17}
\subsubsection{Spectroscopic data}
\label{sec:specdata}

We make use of high-quality elemental abundances for 136701 red giant star
from APOGEE Data Release 17 \citep[APOGEE DR17,][]{2016AN....337..863M, 2022ApJS..259...35A} and {\it Gaia} 
Early Data Release 3 \citep[Gaia EDR3,][]{2016A&A...595A...1G, 2021A&A...649A...1G} to hunt for merger remnants
in the Milky Way stellar disk. The sample was selected to contain only stars with $\sigma_{\pi}/\pi < 0.2$. For quality control of the stellar
parameters and elemental abundances, the following selection criteria
for the spectroscopic data were applied:

\begin{itemize}
\setlength\itemsep{0.02cm}
\item{ \tt{SNREV > 80}}
\item{ \tt{AL\_FE\_FLAG = 0}}
\item{ \tt{TEFF < 6000 K}}
\item{ \tt{TEFF > 4000 K}}
\item{ \tt{LOGG < 2.8}}
\item{ Remove any {\tt VERY\_BRIGHT\_NEIGHBOR} and {\tt PERSIST\_HIGH} bit set in {\tt STARFLAG}}
\item{ Remove any {\tt STAR\_BAD}, {\tt CHI2\_BAD}, {\tt M\_H\_BAD}, and {\tt CHI2\_WARN} bit set in {\tt ASPCAPFLAG}}
\item{ Remove duplicate observations of a single target that have bit 4 of {\tt EXTRATARG} set}
\end{itemize}

The upper effective temperature limit was imposed to remove stars at the hot edge of the main stellar atmospheric model grids used in the APOGEE analysis. For some of these stars, the effective temperatures derived have preferentially discrete values. 
The lower effective temperature limit was imposed to remove cool red giant stars with suspected problems with the derivation of the aluminium and magnesium elemental abundances. 
The sample was further limited to giant stars in order to avoid any possible effects of an over-density of stars
seen in the [Al/Fe] of metal-rich dwarfs stars. For more details on the APOGEE DR17 data quality see
the SDSS DR17 documentation available on the web\footnote{https://www.sdss.org/dr17/} and a detailed discussion of DR16 in \citet{2020AJ....160..120J}.

The dataset has been cleared of all fields and programs targeting known stellar clusters, accreted stellar populations, and stars outside the Milky Way
\citep{2021AJ....162..303S,2021AJ....162..302B}. The full lists of the removed fields and program names are given in Tables \ref{tab:fields} and \ref{tab:progname}, respectively.

\subsubsection{Calculation of kinematic data}

Astrometric measurements from {\it Gaia} EDR3 are available for the APOGEE sample and the RR~Lyrae stars (see Sect. \ref{sec:rrlyr}). We use the more precise APOGEE spectroscopic radial velocity measurements when calculating kinematics for the APOGEE sample. 
Galactic positions and velocities are calculated for all stars using Astropy \citep{astropy:2013, astropy:2018} and {\it galpy} \citep{2015ApJS..216...29B}. { We use the \textit{actionAngleStaeckel} approximation \citep{2013ApJ...779..115B, 2012MNRAS.426.1324B} with the MWPotential14 Milky Way potential \citep{2013ApJ...779..115B}, delta of 0.4, and default values for other parameters.
The mean uncertainties for our sample are 0.030 km s$^{-1}$ in radial velocity, 0.020 mas year$^{-1}$ in RA proper motion, and 0.018 mas year$^{-1}$ in Dec proper motion. We expect the resulting uncertainties on the calculated kinematics to be small, specifically the uncertainty in rotational velocity ($V$), which we use as a selection parameter, is $\sim 10$ km s$^{-1}$.}
The photogeometric distance from \citet{2021AJ....161..147B} is used for all stars.

\subsubsection{Identification of the population}

\begin{figure}
    \centering
    \includegraphics[clip, trim=0cm 0cm 1cm 1cm, angle=0, width=0.49 \textwidth]{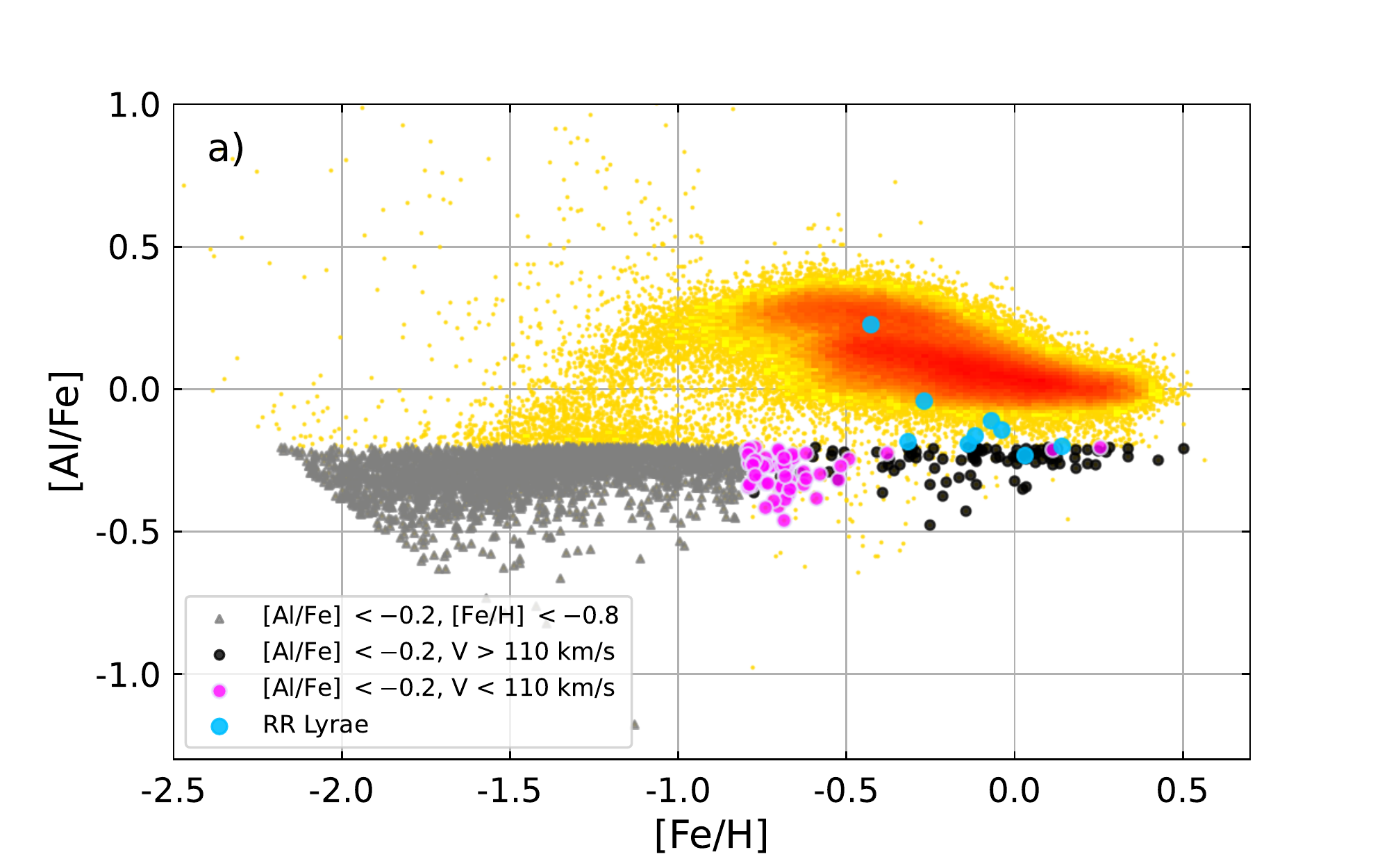}
    \includegraphics[clip, trim=0cm 0cm 1cm 1cm, angle=0, width=0.49 \textwidth]{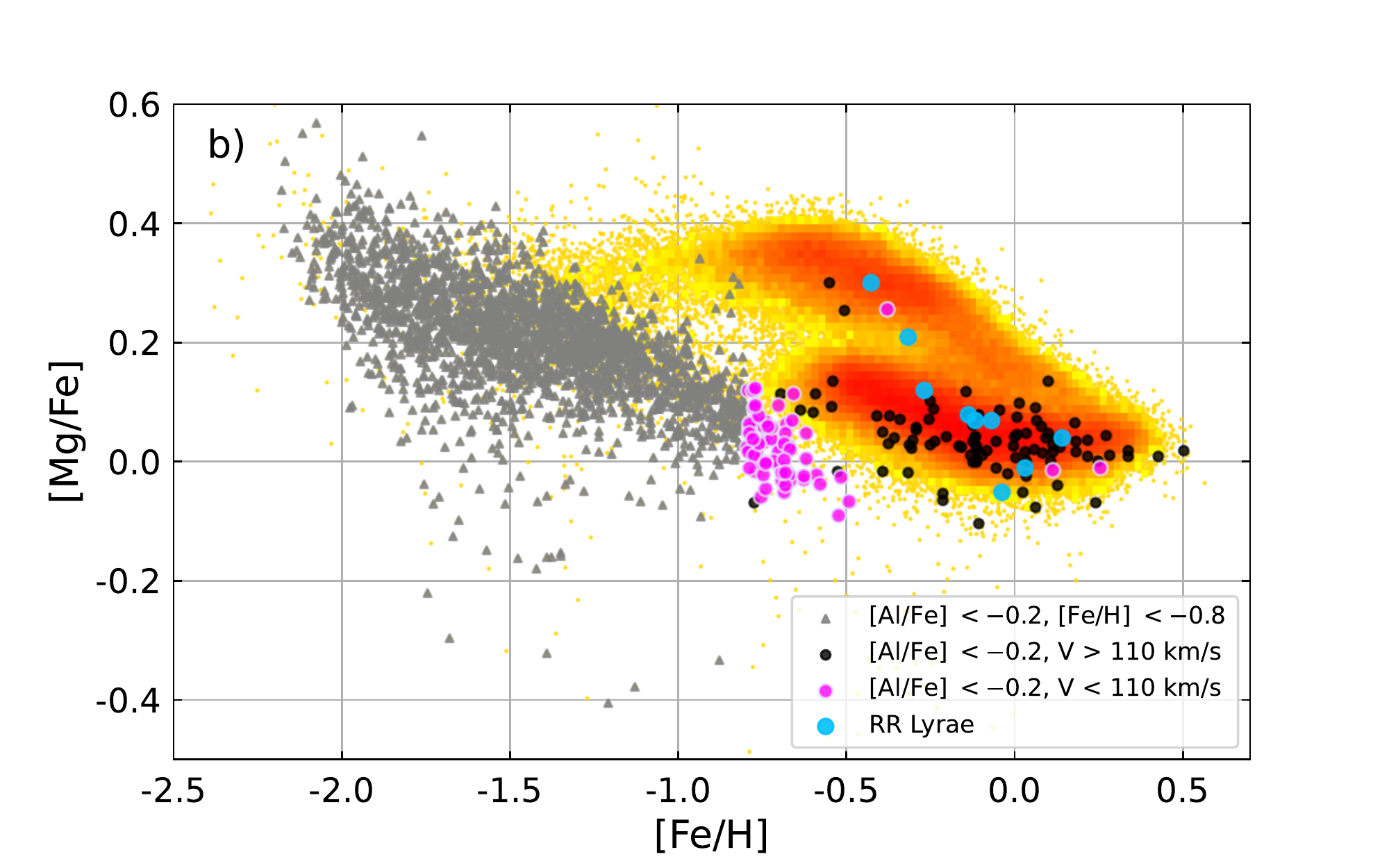}
    \includegraphics[clip, trim=0cm 0cm 1cm 1cm, angle=0, width=0.49 \textwidth]{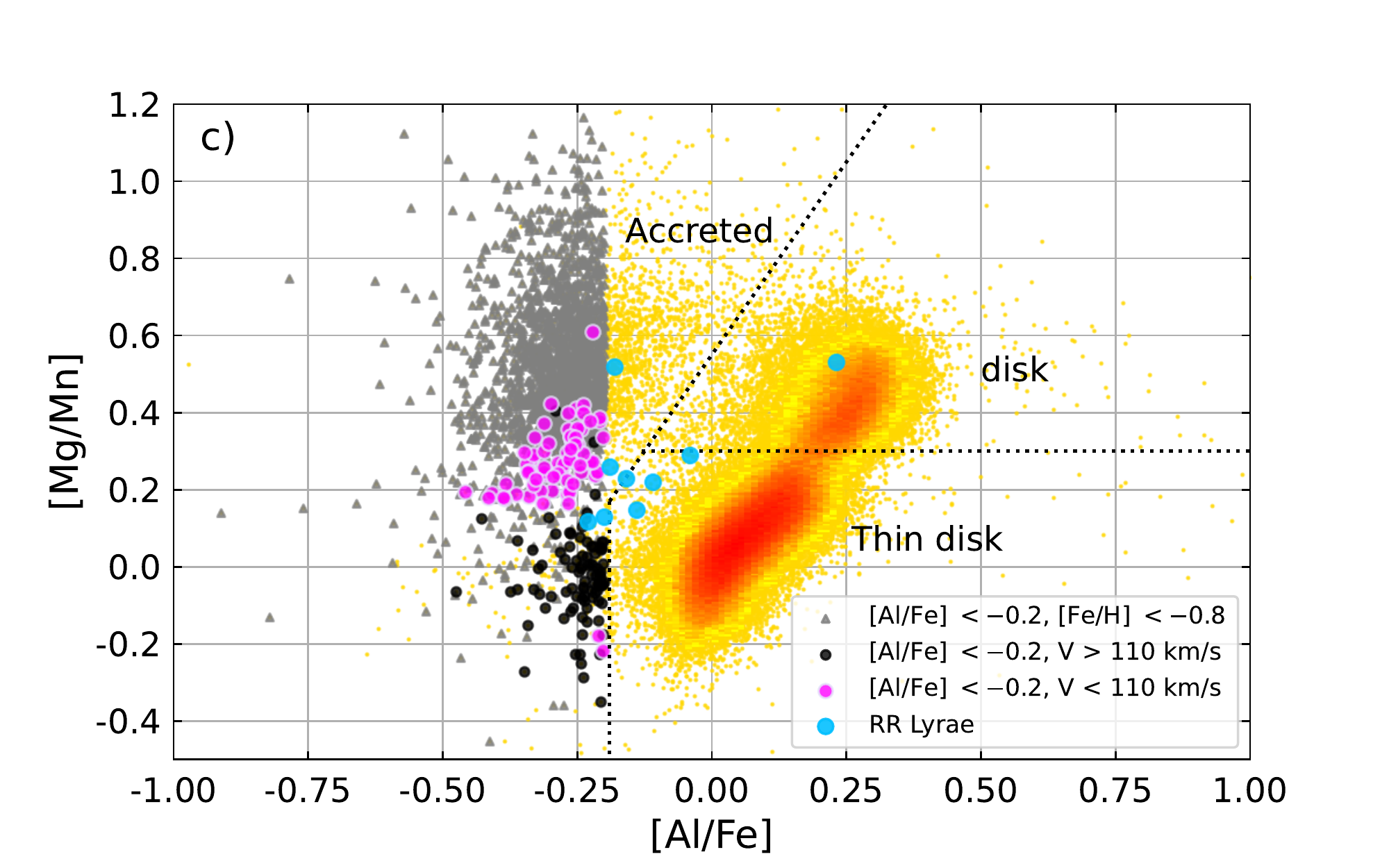}
    \caption{a) [Al/Fe] as a function of
      [Fe/H]. This figure illustrates the defining cut in [Al/Fe].
      Symbols as indicated in the inset. A discussion of the different
      cuts in [Al/Fe] used for the APOGEE DR17 and RR~Lyrae samples
      can be found in Sect.~\ref{sect:rrlyrpop}. 
      b) [Mg/Fe] as a function of [Fe/H]. Symbols as indicated in the inset.
      c) [Mg/Mn] as a function of [Al/Fe]. Symbols as indicated in the inset.
      Note the different scales on each axis.}
    \label{fig:abun}
\end{figure}

\begin{figure}
    \centering
        \includegraphics[clip, trim=0cm 0.5cm 1cm 1.5cm, angle=0, width=0.49 \textwidth]{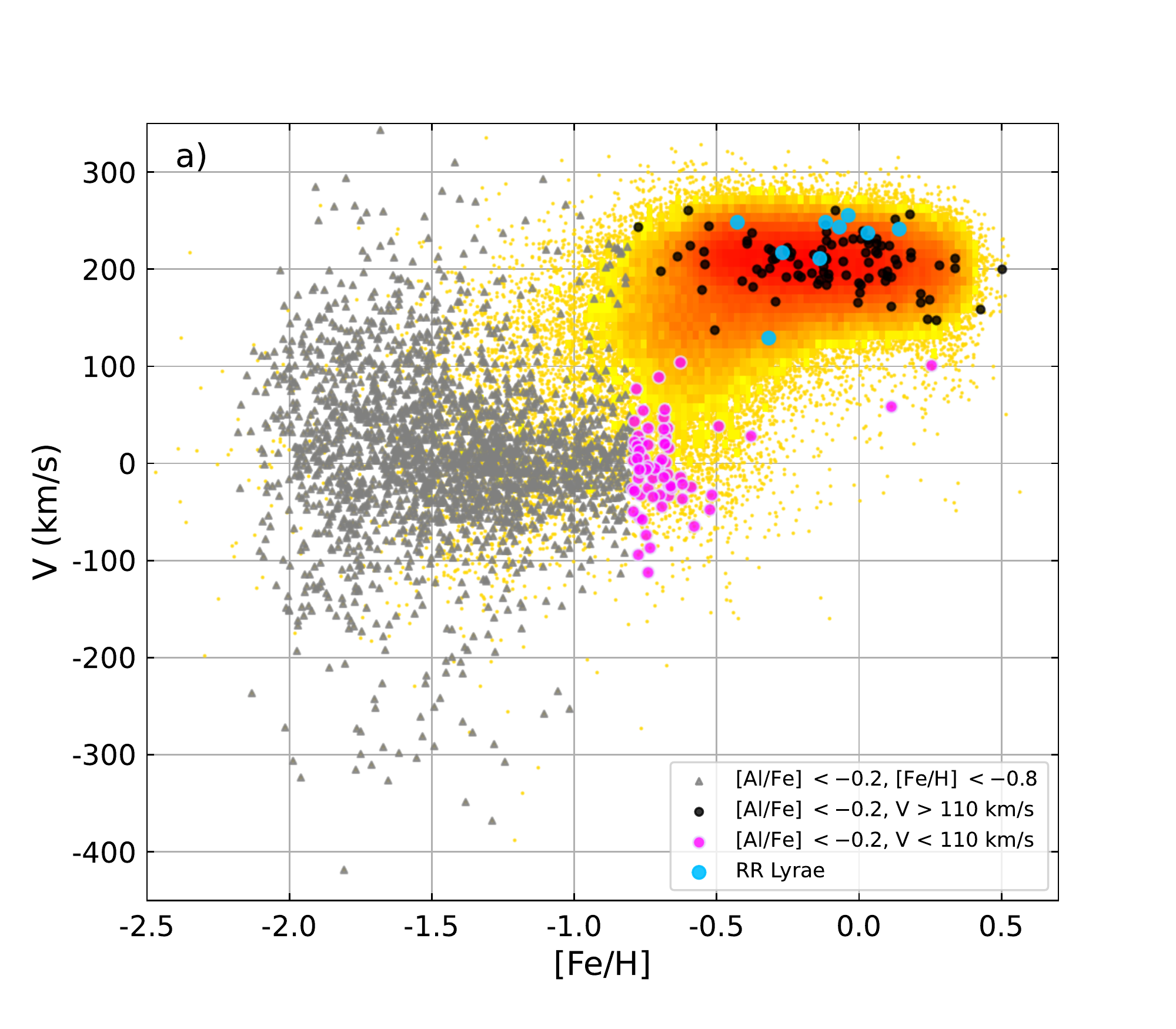}
    \includegraphics[clip, trim=0cm 0.5cm 1cm 1.5cm, angle=0, width=0.49 \textwidth]{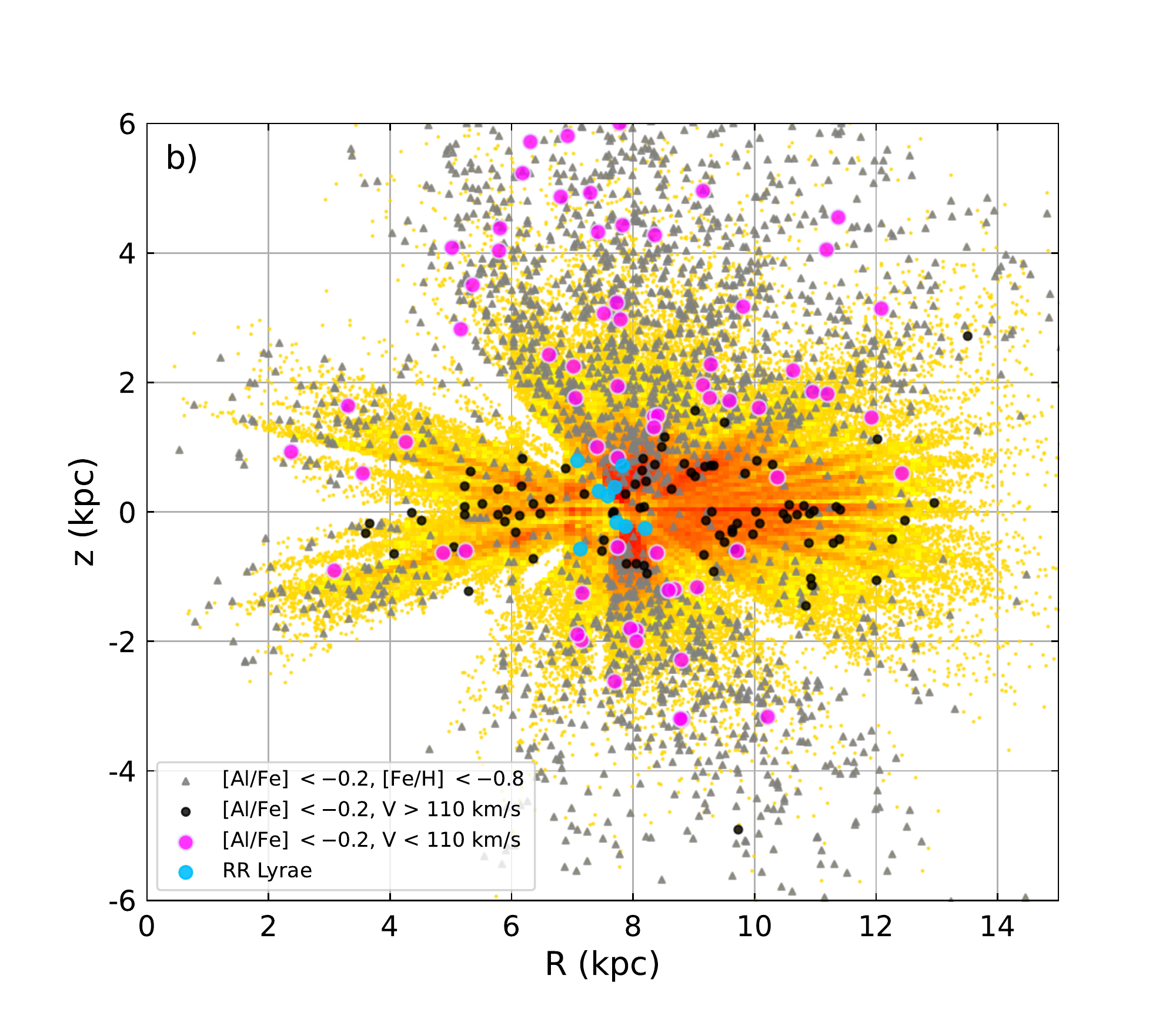}
    \caption{a) Rotational velocity ($V$) as a
      function of [Fe/H]. Symbols as indicated in the inset.
      b) Positions of the stars in the Milky
      Way, vertical position ($z$) as a function of distance from the
      Galactic center ($R$). Symbols as indicated in the inset.}
    \label{fig:kinpos}
\end{figure}


\begin{figure*} 
    \centering
    \includegraphics[clip,width=0.75 \hsize,angle=0,trim=2.75cm 2.2cm 2.8cm 2.5cm]{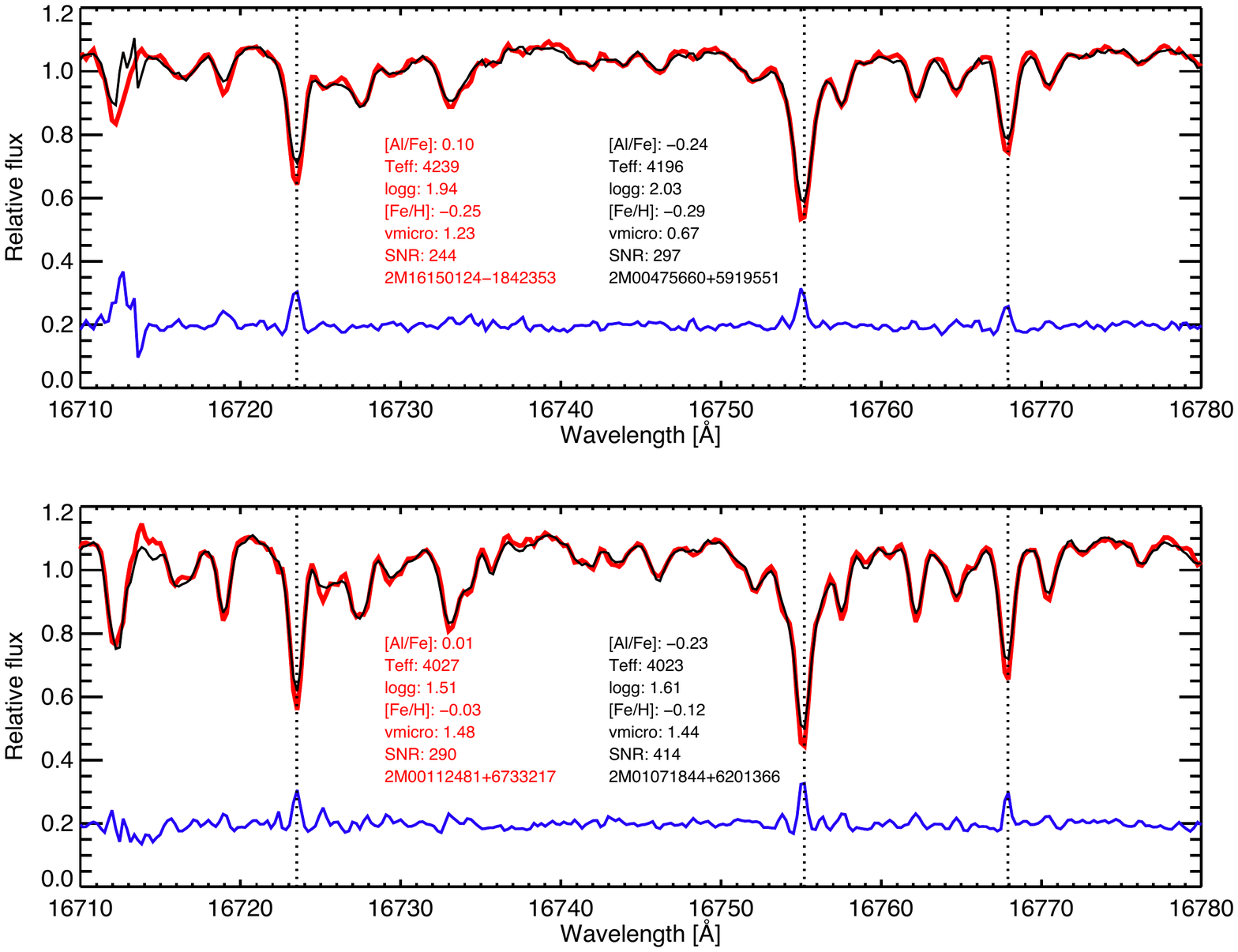}
    \caption{Observed spectra for two pairs of stars illustrating the reality that the aluminum lines are weaker in the stars with low-[Al/Fe]. In each panel we show two stars with similar stellar parameters. The red spectra have [Al/Fe] values typical of the stellar disk, whilst the black spectra have low [Al/Fe] values. The blue line shows the spectrum of the low-[Al/Fe] star divided by that of the [Al/Fe]-normal star, emphasizing the difference in the strength of the aluminum lines, indicated by the vertical dotted lines, that are otherwise remarkably similar.}
    \label{fig:spectra}
\end{figure*}

{ Through the following investigation, we have identified ``disk'' stars to have [Fe/H]~$> -0.8$ and $V > 110$~km~s$^{-1}$.} 
We first looked for stars with low-[Al/Fe] and disk-like
metallicities ([Fe/H] $> -0.8$). Fig.~\ref{fig:abun}a shows that such stars exist { and are identified as black points}. We define a low-[Al/Fe] star to have $\rm [Al/Fe]<-0.2$. This cut is
somewhat arbitrary but is informed by the locus of accreted stars at lower metallicities found in previous studies 
\citep[][]{2015MNRAS.453..758H, 2020MNRAS.493.5195D, 2021MNRAS.508.1489F}.

The properties of these stars are further explored in
Fig.~\ref{fig:abun}b,c and Fig.~\ref{fig:kinpos}. { In these figures, the cleaned APOGEE DR17 data is shown as the orange-scale density and yellow points; low-[Al/Fe], low metallicity stars ([Al/Fe] $< -0.2$, [Fe/H] $< -0.8$) are shown as gray triangles; low-[Al/Fe] disk-like stars are shown as black points; low-[Al/Fe] stars with disk-like metallicity but halo-like velocity are shown as pink points; and the RR~Lyrae sample is shown as blue points.}

We see that stars identified as ``disk-stars'' using a cut in the $V$ - metallicity plane ({ $V > 110$ km s$^{-1}$}, Fig.~\ref{fig:kinpos}a) follow the typical thin disk trend for [Mg/Fe] (Fig.~\ref{fig:abun}b) and that they are essentially confined to the plane of the Milky Way (Fig.~\ref{fig:kinpos}b). 
{ These stars are also separate from the halo-like accreted stars in [Mg/Mn]-[Al/Fe] space, Fig.~\ref{fig:abun}c. Known accreted halo stellar populations, such as {\it Gaia}-Sausage-Enceladus, have been found to occupy a distinct region of this elemental abundance space \citep[e.g.][]{2020MNRAS.493.5195D}. The dotted lines in Fig.~\ref{fig:abun}c show the approximate separation of accreted, thick disk, and thin disk stars in this space. We note that previous studies have included only the halo-like stars with [Mg/Mn] $> 0.1$ as the accreted region.
These chemical, spatial, and kinematics properties give us confidence
that we have identified a sample of stars that are kinematically integrated with the disk, have anomalously low aluminum abundances, but are chemically distinct from previously identified accreted halo populations.} 

\subsubsection{Checking that the low-[Al/Fe] abundances are real}
\label{sect:spec}

To check if the low-[Al/Fe] measurements for stars with disk-like metallicities are robust
and not the product of problems in the elemental abundance determination we
performed an ocular inspection of the stellar spectra of all the
low-[Al/Fe] stars with [Fe/H] $> -0.8$. We compared each low-[Al/Fe] spectrum with the spectrum of
another star with the same stellar atmospheric parameters but with ordinary
[Al/Fe] values. Fig.~\ref{fig:spectra} shows two examples of these
comparisons. They show that indeed the aluminum lines in the low-[Al/Fe]
stars, marked by the vertical dotted lines, are shallower than in the stars with ordinary [Al/Fe] values.
This comparison also shows that our finding is not the result of NLTE
effects that may not have been accounted for by the APOGEE DR17 analysis.
The direct comparison of stars with the same stellar parameters shows
the difference to be real. 

\begin{figure}
    \centering
    \includegraphics[clip, trim=0.5cm 0.5cm 2cm 1.5cm, angle=0, width=0.49 \textwidth]{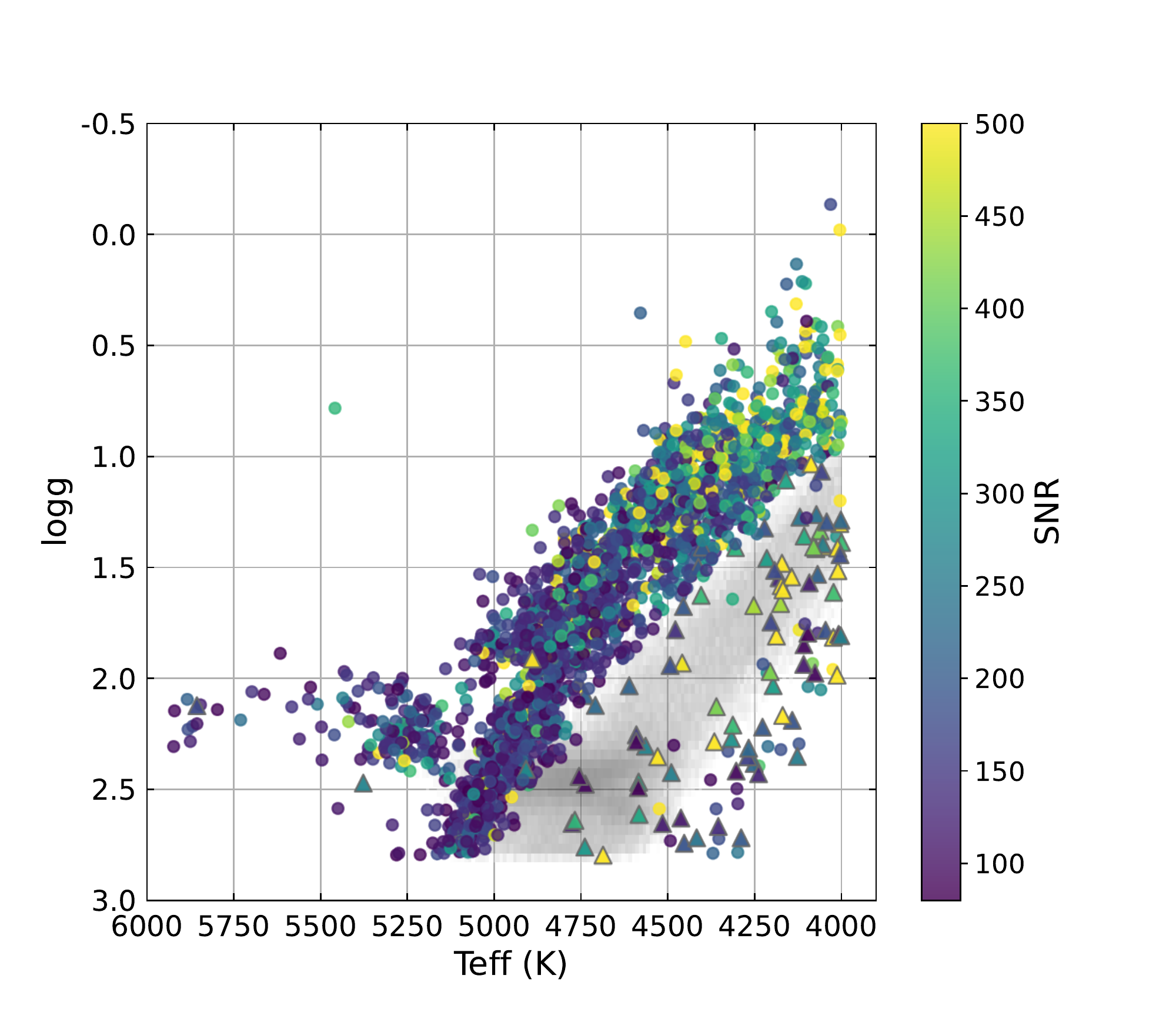}
    \caption{The low-[Al/Fe] APOGEE stars in the $\log g - T_{\rm eff}$ plane. Stars with disk-like
    kinematic properties are identified by triangles and stars with halo-like properties with points. Those stars shown with [Fe/H] $> -0.8$ have passed spectral inspection. The distribution of the full APOGEE sample is shown in gray for comparison.
    }
    \label{fig:hr}
\end{figure}

In Fig.~\ref{fig:hr} we show the APOGEE stars identified as low-[Al/Fe] stars ([Al/Fe] $<-0.2$) in the $\log g - T_{\rm eff}$ plane with the full APOGEE sample in gray for comparison. It is clear that the more metal-rich disk-like stars have cooler temperatures whilst the halo like stars are warmer, indicating a color-difference such that the halo is blue and the disk red. The color-coding indicates the signal-to-noise ratio (SNR) in the APOGEE spectra used for the abundance analysis. As can be seen, the spectra have good SNR.

We  have also inspected the derived abundances as a function of $T_{\rm eff}$, $\log g$, SNR, and microturbulence and find no obvious  trends for any of the APOGEE  subsamples used.

To summarize, the spectra for the low-[Al/Fe] stars are of good quality and we have no reason to believe that the results are the result of poor quality spectra. Combined with the visual inspection of all the metal-rich low-[Al/Fe] spectra, we have weeded out all spectra that were not reliable and what remains is a clean sample.

\subsection{Sample of RR~Lyrae variables}
\label{sec:rrlyr}

\begin{table*}
\centering
\caption{Summary of available data for the nine known RR~Lyrae variables with
measured aluminum abundances. Elemental abundances are from \citet{2013RAA....13.1307L} (Liu) and \citet{2019ARep...63..203M} (Marsakov).
Masses are taken from \citep{2019ARep...63..203M}. We indicate which stars
have \textit{Gaia} radial velocities available. For those without \textit{Gaia} radial velocities we use values from \citet{1994AJ....108.1016L}. The last column indicates if we identify the star as an accreted star based on elemental abundance criteria. }
\label{tab:rrlyr}
\begin{tabular}{l c c c c c c c c c c}
\hline
Star & \multicolumn{2}{c}{[Fe/H]} &  [Mg/Fe] & [$\alpha$/Fe] &       [Al/Fe]   &  [Mn/Fe]  &   Mass [M$_{\sun}$] & \textit{Gaia} RV & Layden RV &  Accr.\\
& Liu & Marsakov & Liu & Marsakov & Liu & Liu & Marsakov & & & \\
   \hline \hline
AA~Aql & --0.32&	--0.32&	{\color{white}--}0.21 &	{\color{white}--}0.18 &	--0.18&	--0.31&	0.56 &  & \checkmark &  \checkmark\\
CN~Lyr & --0.04&	--0.04&	--0.05&	--0.01&	--0.14&	--0.20&	0.54 & \checkmark & & \checkmark \\
DM~Cyg & {\color{white}--}0.03&	{\color{white}--}0.03 &	--0.01&	--0.06&	--0.23&	--0.13&	0.54& \checkmark & & \checkmark\\
DX~Del & --0.14&	--0.31&	{\color{white}--}0.08&	--0.02&	--0.19&	--0.18&	0.54& \checkmark & & \checkmark \\
KX~Lyr & --0.27&	--0.42&	{\color{white}--}0.12&	{\color{white}--}0.09&	--0.04&	--0.17&	0.58& \checkmark & & \\
RS~Boo & --0.12&	--0.21&	{\color{white}--}0.07&	{\color{white}--}0.03&	--0.16&	--0.16&	0.53& \checkmark &  & \checkmark\\
SW~And & --0.07&	--0.22&	{\color{white}--}0.07&	{\color{white}--}0.00&	--0.11&	--0.15&	0.53 &  & \checkmark& (\checkmark) \\ 
TV~Lib & --0.43&	--0.43&	{\color{white}--}0.30&	{\color{white}--}0.29&	{\color{white}--}0.23&	--0.23&	0.54 &  & \checkmark &\\
V445~Oph & {\color{white}--}0.14& {\color{white}--}0.11&	{\color{white}--}0.04&	--0.05&	--0.20& 	--0.09&	0.52 & \checkmark & & \checkmark\\
 \hline
 \end{tabular}
\end{table*}

\subsubsection{Spectroscopic data and population identification}
\label{sect:rrlyrpop}

RR Lyrae variable stars are known to be old and are often used to characterize accreted populations in the Milky Way \citep{2013MNRAS.435.3206D}. \citet{2013RAA....13.1307L} identify a sample of 23 RR Lyrae stars belonging to the disk of the Milky Way. To complement the sample of low-[Al/Fe] APOGEE disk stars, we make use of this RR Lyrae sample. We select the nine stars with aluminum measurements to use in our analysis. Table~\ref{tab:rrlyr} lists the sub-sample and compiled parameters used in this work. The [Fe/H], [Mg/Fe], [Al/Fe], and [Mn/Fe] abundance ratios are derived by \citet{2013RAA....13.1307L}. For comparison, we include [Fe/H] and [$\alpha$/Fe] values compiled in \citet{2019ARep...63..203M}. Four of the stars have identical [Fe/H] in the two studies. In those cases, the magnesium and $\alpha$ abundances (calculated as the average of magnesium, calcium, silicon, and titanium) are also almost identical. For four stars there is a large difference in the [Fe/H] values in the two studies. In this work we use exclusively the data from \citet{2013RAA....13.1307L}, which presents data from an homogeneous analysis. 

To investigate if any of the selected RR~Lyrae variables are accreted disk stars we make use of two elemental abundance diagnostics: 1) A cut in [Al/Fe] at $-0.14$~dex, Fig.~\ref{fig:abun}a, and 2) the diagnostic plot in the [Mg/Mn] - [Al/Fe] plane, Fig.~\ref{fig:abun}c, proposed by \citet{2015MNRAS.453..758H}. The exact cut in [Al/Fe] must remain arbitrary, in particular since we have no way of knowing if the APOGEE DR17 and the data from \citet{2013RAA....13.1307L} are on the same absolute scale. We have chosen to set the cut at $-0.14$~dex for the RR~Lyrae variables. Accreted stellar populations have been empirically found to occupy a different region of the [Mg/Mn] - [Al/Fe] abundance plane than Milky Way stellar populations \citep{2020MNRAS.493.5195D}. There is some theoretical motivation for this \citep[see][]{2015MNRAS.453..758H} and initial chemical evolution models confirm a different evolutionary sequence through this space for a population similar to the {\it Gaia}-Sausage-Enceladus \citep{2021MNRAS.500.1385H}.

It should be noted that some elemental abundances can change in RR~Lyrae variables as a function of the pulsation phase \citep{2011ApJS..197...29F,2013RAA....13.1307L}. Examining Table~6 in \citet{2013RAA....13.1307L} we see that some abundances, for example manganese and magnesium, remain stable throughout the phases whilst others may have some changes, notably aluminum. In the two stars with [Al/Fe] measurements at different phases, the variation in [Al/Fe] is as large as $0.15$~dex. Our understanding of this effect is that the change in abundance, if anything, makes it harder to identify the accreted stars (i.e. they may have higher [Al/Fe] in one phase). Thus our identification of low-[Al/Fe] variable stars should be conservative. 

A few of the RR~Lyrae in Table~\ref{tab:rrlyr} are not included in our sample of accreted stars. Of the nine stars one, TV~Lib, shows thick disk elemental abundances with elevated [Mg/Fe] and [Al/Fe]. We do not associate KX~Lyr with the accreted sample as it has approximately solar [Al/Fe]. The [Al/Fe] value for SW~And falls just short of our (arbitrary) cut at $-0.14$~dex, we therefore include it in our accreted sample, but use caution when drawing conclusions that may be influenced by its presence.  AA~Aql, has low [Al/Fe] but higher [Mg/Fe]. When examined in the [Mg/Mn]-[Al/Fe] plane the star clearly falls in the accreted halo region (see Fig.~\ref{fig:abun}c). This star also has a rotational velocity ($V$) more associated with thick disk/halo kinematics than with thin disk kinematics (Table~\ref{tab:rrlyr}). The remaining five stars have a high [Fe/H], centered near solar values and all have low-[Al/Fe] and fall in the disk region in Fig.~\ref{fig:kinpos}a. Our final classification of accreted and non-accreted for the nine RR~Lyrae stars can be found in Table~\ref{tab:rrlyr}. In total we are able to identify six stars with low-[Al/Fe], suggesting they could be accreted, with a seventh very close to the cut in [Al/Fe]. Six of the seven have kinematics consistent with the thin disk.

\subsubsection{Estimating masses of RR~Lyrae variable stars}
\label{sect:rrlyrmass}

RR~Lyrae variables are stars that have evolved from the main
sequence, along the giant branch and have descended onto the
horizontal branch and are currently crossing the instability strip
\citep[a good explanation of this from the point of view of population
synthesis is given in][]{2020A&A...641A..96S}. If we know the mass of
an RR~Lyrae variable and how much mass it loses along the red giant
branch then we can infer its mass on the main-sequence. This mass
relates directly to the age of the stellar population of which that the
RR~Lyrae variable is a member.

Mass-loss along the red giant branch is difficult to measure, but
recent studies enabled by a large data-set for globular clusters and
dwarf spheroidal galaxies have been able to show that the mass-loss
along the red giant branch measured empirically is dramatically
different from many of the older theoretical estimates
\citep{2020MNRAS.498.5745T,2020A&A...641A..96S}.


Current mass estimates are available for
all stars from \citet{2019ARep...63..203M}. We note that the [Fe/H]
values for two of the accreted stars are not the same in the study
that derived the elemental abundances and the compilation used to
derive the stellar masses (compare Table~\ref{tab:rrlyr}). However,
the differences are not large enough to change the masses of these
stars drastically and certainly not large enough to make the stars
much heavier \citep{2019ARep...63..203M}.
	
The current masses for the  RR~Lyrae variables in our sample are all
around 0.5 to 0.6~M$_{\odot}$ (Table~\ref{tab:rrlyr}). Using the
recent mass-loss laws by \citet{2020MNRAS.498.5745T} we estimate the
mass-loss for our low-[Al/Fe] RR~Lyrae variables and find that their
turn-off masses are not in excess of 0.85~M$_{\odot}$. Such masses
indicate an old stellar population, approximately 10-12 Gyr.

\subsection{Kinematic data}
\label{sec:kin}

\begin{figure}
    \centering
    \includegraphics[width=8.5cm]{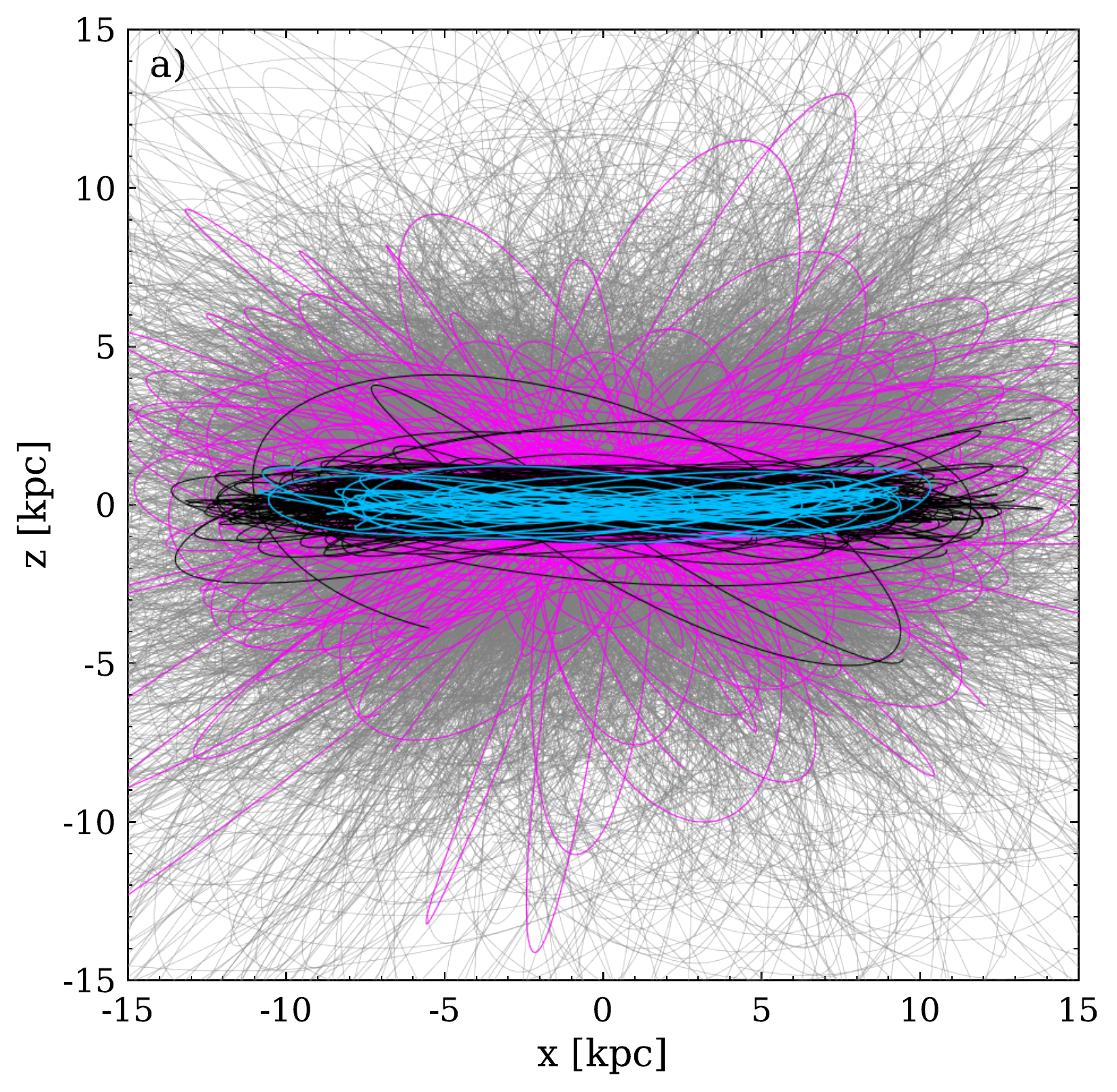}
    \includegraphics[width=8.5cm]{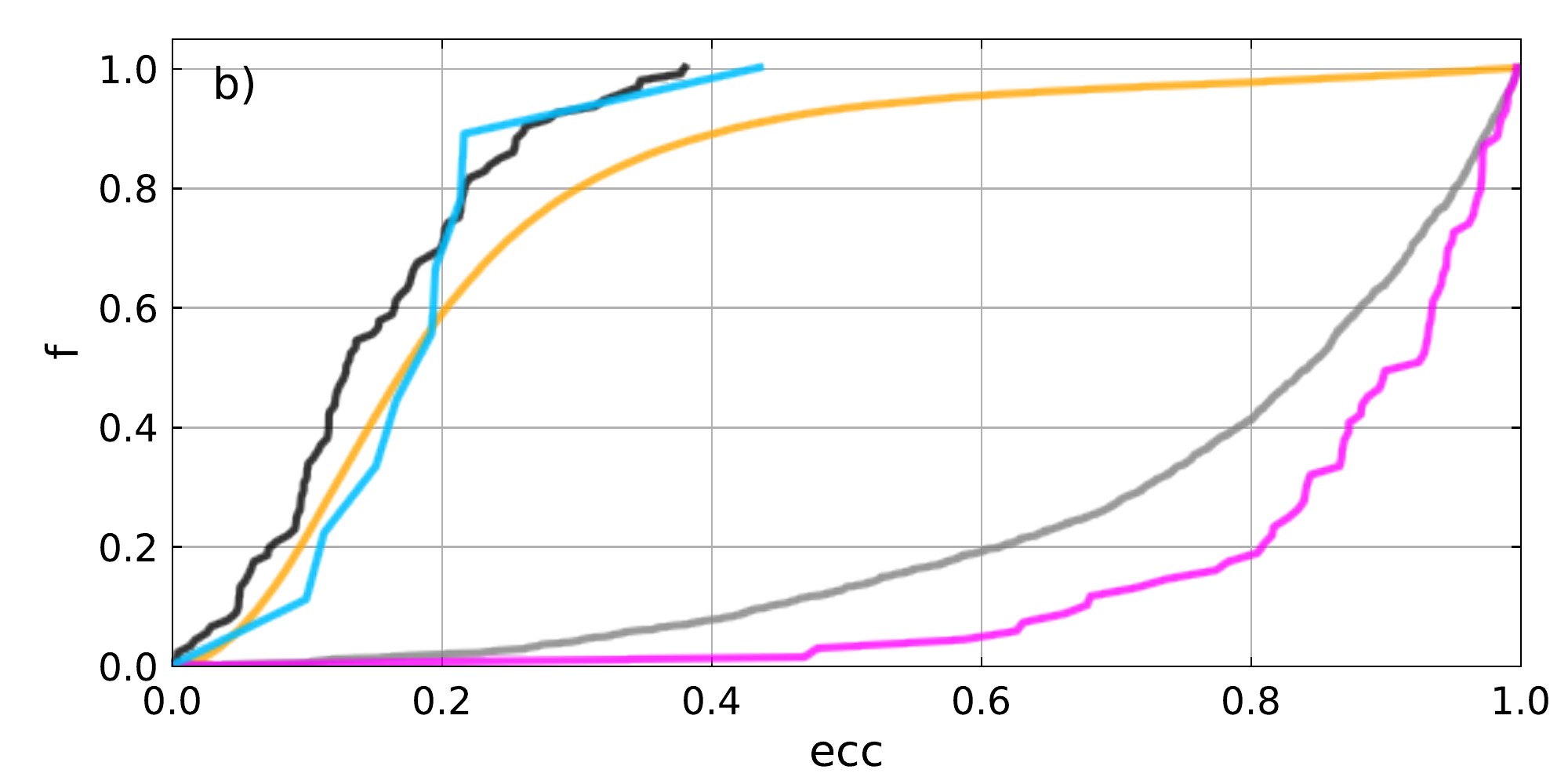}
    \includegraphics[width=8.5cm]{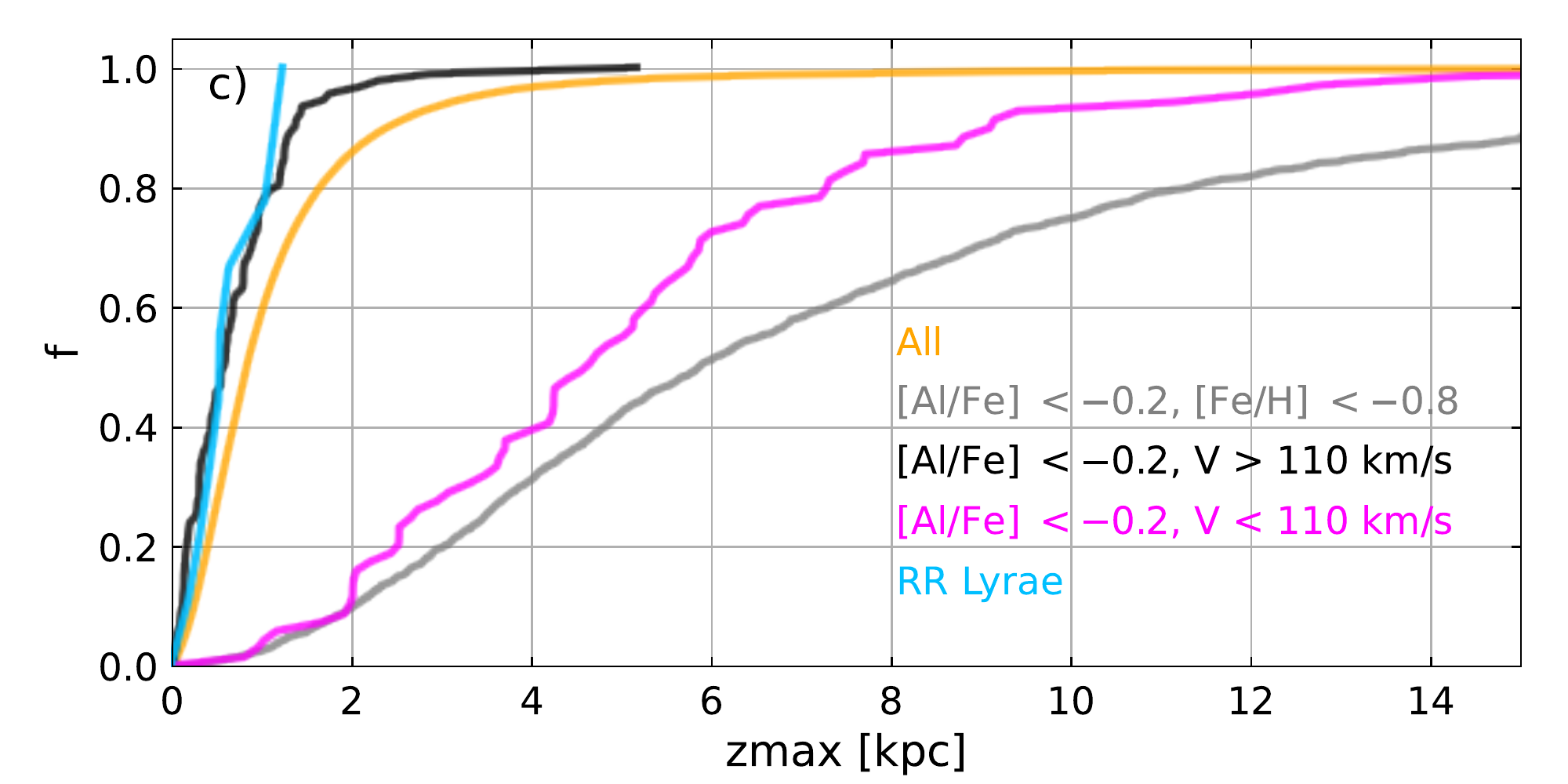}
    \caption{a) Orbits for the full sample of selected APOGEE DR17 red
      giant branch stars with [Al/Fe] < -0.2, [Fe/H] < -0.8 (gray lines), 
      [Al/Fe] < -0.2, V < 110 (magenta lines)
      and [Al/Fe] < -0.2, V > 110 (black lines),
      and RR~Lyrae variables (blue lines).
      b) Cumulative histogram of orbital eccentricities.
      c) Cumulative histogram of orbit z max values.}
    \label{fig:orbits}
\end{figure}



All nine RR~Lyrae stars have astrometric measurements from {\it Gaia} EDR3 are available. Six of the RR~Lyrae stars have \textit{Gaia} RVS measurements. { We prefer to use \textit{Gaia} RV measurements when possible to keep the dataset homogeneous and because multiple measurements, such as done by \textit{Gaia}, reduces the uncertainty in the mean RV of these pulsating stars. The RV uncertainty is 4-5 km s$^{-1}$. For the remaining three RR~Lyrae, the radial velocities are taken from new data table of \citet{1994AJ....108.1016L}, which uses at least two measurements. The RV uncertainty is 15-20 km s$^{-1}$. The distance for all RR~Lyrae is calculated from the \textit{Gaia} parallax measurement using the astropy Distance function.}

Fig.~\ref{fig:kinpos}b shows the Galactic positions of the low-[Al/Fe] stars in comparison to the full sample. We see that the stars with disk-like kinematics (Fig.~\ref{fig:kinpos}a) are spread across the whole disk, but are mainly confined to the plane. Of the APOGEE stars with disk-like kinematics, 12 stars out of 92 low-[Al/Fe] have current vertical positions larger than 1 kpc from the Galactic plane. For low-[Al/Fe] stars with halo kinematics these numbers are 1873 stars out of 2077 stars.

The orbits for the low-[Al/Fe] stars were integrated using {\it galpy} in the MWPotential14 Milky Way potential \citep{2013ApJ...779..115B} for 500 Myr. 
The orbital trajectories of the low-[Al/Fe] stars are shown in Fig.~\ref{fig:orbits}a with disk stars (blue) and halo stars (black) compared to the RR~Lyrae stars (red). Fig.~\ref{fig:orbits} also shows cumulative histograms of the orbital eccentricity (b) and z max (c) of the APOGEE sub-samples as well as the RR~Lyrae. The orbits of the low-[Al/Fe] stars identified as belonging to the cool disk are confined to the Galactic plane, further strengthening their association with the disk. These stars are seen to be on mainly circular orbits with no orbits reaching inside about $R_{\mbox{Gal}} = 4$~kpc.

\section{Results and discussion}

\subsection{Accreted disk stars identified in APOGEE DR17}

Fig.~\ref{fig:abun} shows three plots of elemental abundance data for our sample as well as the RR~Lyrae stars with likely accreted origins. Panel a in Fig.~\ref{fig:abun} shows [Al/Fe] as a function of [Fe/H]. This plot is used to define the low-[Al/Fe] stars, which we propose may have accreted origins. Empirically, several dwarf galaxies and known accreted stellar populations in the Milky Way have low [Al/Fe] abundances. This is nicely demonstrated in \citet[][see their Fig.~5]{2021ApJ...923..172H}, which shows APOGEE elemental abundances of the Large Magellanic Cloud, Small Magellanic Cloud, \textit{Gaia}-Sausage-Enceladus accreted population, Sagittarius dwarf galaxy, and Fornax dwarf galaxy in comparison to Milky Way disk stars with a similar range of effective temperature and surface gravity. These five ex situ stellar populations all have [Al/Fe] abundance ratios well below the Milky Way at a given [Fe/H]. Of particular relevance to our study is the Sagittarius dwarf galaxy, which maintains a low [Al/Fe], $\sim-0.4$ dex, even at metallicities approaching solar values. This supports the possibility of low-[Al/Fe] stars forming at high [Fe/H] in a dwarf galaxy environment.


Beyond using [Al/Fe] alone as an indicator of ex situ origins, the [Mg/Mn] vs [Al/Fe] space has been identified as holding high diagnostic power in this effort \citep{2020MNRAS.493.5195D, 2015MNRAS.453..758H, 2021MNRAS.500.1385H}. 
Although the cause of the low [Al/Fe] abundance ratios found in ex situ populations is not well understood, the theoretical motivation for using the [Mg/Mn] vs [Al/Fe] elemental abundance space as detailed by \citet{2015MNRAS.453..758H} and \citet{2020MNRAS.493.5195D} can inform our discussion.
We therefore seek some understanding in the known formation channels of aluminum, magnesium, and manganese (see \citet{2020ApJ...900..179K} for a detailed modeling of element production in a Milky Way environment). Magnesium is produced primarily in type II supernovae (SNII) and in a small amount by Asymptotic Giant Branch (AGB) stars. Mangenese is primarily produced in type Ia supernovae (SNIa), with small contributions from SNII and AGB stars. Aluminum is produced primarily in SNII with small contributions from AGB stars. 
In contrast to the low-[Al/Fe] disk stars presented in this work, the five dwarf galaxy or accreted populations discussed in \citet{2021ApJ...923..172H} maintain [Mg/Fe] abundance ratios below that of the Milky Way disk at any given [Fe/H].
\citet{2013ApJ...778..149M} conclude that the low [$\alpha$/Fe] abundance ratios in the Sagittarius dwarf galaxies are an indication of a lack of high-mass stars or a top-light initial mass function. They argue that the low [Al/Fe] supports this finding as aluminum is formed during hydrostatic carbon and neon burning in massive stars \citet{1995ApJS..101..181W}.

Panel b of Fig.~\ref{fig:abun} shows [Mg/Fe] as a function of [Fe/H]. The low-[Al/Fe] stars with [Fe/H] less than about -0.6 dex essentially all fall on the lower [Mg/Fe] sequence, which has been associated with accreted populations generally \citep{2010A&A...511L..10N, 2018ApJ...852...49H} and specifically the \textit{Gaia}-Enceladus-Sausage and Sequoia merger events \citep{2020ARA&A..58..205H, 2020RNAAS...4..246M, 2021MNRAS.508.1489F}. At higher [Fe/H] the low-[Al/Fe] stars either fall on the thin disk sequence or below the main sample. Only two APOGEE stars and one RR~Lyrae fall on what is commonly referred to as the thick disk trend ([Mg/Fe] > 0.2 dex). \citet{2021ApJ...923..172H} find that the Sagittarius population reaches similarly high [Fe/H] and low [Al/Fe], but does not have [Mg/Fe] consistent with the Milky Way thin disk.

Fig.~\ref{fig:kinpos} shows [Fe/H] as a function of the tangential velocity ($V$) for all low-[Al/Fe] stars. The low-[Al/Fe] stars fall into two groups; one with typical disk-like kinematics and one with halo-like kinematics. The latter are centered around $V=0$ km s$^{-1}$, in fact they are somewhat retrograde, consistent with findings of the kinematics of the major remnants in the Milky Way halo \citep{2020ARA&A..58..205H, 2020MNRAS.497..109F}. We have calculated the orbits of the low-[Al/Fe] stars and find that the stars with [Fe/H] greater than about $-0.8$ dex all have orbits that are kinematically cool, i.e. confined to the Galactic plane, and with mainly circular orbits, see Fig.~\ref{fig:orbits}.

{ The fraction of accreted stars sitting in the disk is a very  difficult parameter to estimate because of selection effects in the APOGEE survey and our selection of this particular dataset. In addition, although we have identified a population of stars that are likely accreted based on their low-[Al/Fe], there are an unknown number of merger events contributing to this population. Given the small number of stars we have robustly identified as low-[Al/Fe] and disk-like out of the full APOGEE sample, the fraction of accreted stars in the Milky Way disk that can be identified through [Al/Fe] is likely very small. 

To roughly estimate this fraction, we consider only stars with $6 < R < 11$ kpc and $|z| < 0.5$ kpc in order to reduce possible spatial biases from the APOGEE lines of sight. Within this spatial region, we limit our estimate to stars with [Fe/H] $> -0.8$ and $V > 110$ km s$^{-1}$, as was required of the low-[Al/Fe] disk stars. The fraction of these stars that have [Al/Fe] $< -0.2$ is $0.073$\%. We tested how this fraction changes as a function of galactocentric radius and [Fe/H] as well as effects of removing any $z$ constraint. The resulting low-[Al/Fe] fraction varies between $0.11$\% and $0.046$\%, with no continuous trends. We stress that these fractions are calculated with very small numbers of low-[Al/Fe] stars and so are very uncertain. The conclude that the stars we identify as accreted disk stars represent approximately $0.07$\% of the APOGEE disk sample.}

\subsection{Age dating the accreted component and the time of the merging}


Our main sample consists of red giant branch stars. Such stars are difficult to determine an age for without access to asteroseismology \citep{2021NatAs...5..640M}.
Another possibility to obtain an age for the newly identified population would be to find stars associated with this population but currently in an evolutionary stage that allow for them to be dated \citep{2010ARA&A..48..581S}. 
{ Instead of directly calculating ages for the red giant stars in our main sample, we have found a sample of intrinsically old stars that are kinematically and chemically associated with the low-[Al/Fe] disk giants.}

Recent studies have found  RR~Lyrae variables with kinematics typical of the cool thin disk and metallicities ranging from a tenth of that of the sun to solar values \citep{2019ARep...63..203M, 2020MNRAS.492.3408P, 2021MNRAS.502.5686I}. In Sect.~\ref{sect:rrlyrpop} we describe the construction of a small sample of RR~Lyrae variables that are associated with the kinematic thin disk and have elemental abundances available in the literature. Some of these stars show the low-[Al/Fe] characteristic for our disk-like accreted sample. 

In Sect.~\ref{sect:rrlyrmass} we derive the masses at turn-off for the RR~Lyrae variables. We find that they all have masses consistent with an old stellar population with an age around 10~Gyr. With an extreme mass-loss of 0.5~M$_{\odot}$
the stars would be somewhat younger, but still several
billion years old \citep{1997ApJ...479..279B}. We should expect fewer metal-rich RR~Lyrae
variables than metal-poor as the mass-range for when a star is in the
instability strip is dependent on metallicity, being very small for the metal-rich stars 
and requiring significant mass-loss, perhaps as
extreme as 0.5~M$_{\odot}$  to become a variable star
\citep[see][]{1976ApJ...207..201T}.



By using kinematic data and elemental abundances we have been able to associate stars of known ages { (i.e. our RR~Lyrae sample) with our newly identified accreted stellar population (i.e. our disk-like, low-[Al/Fe] sample).} These RR~Lyrae stars are all old. This implies that the stellar population that was accreted formed its stars before that time.



Another way to age-date a stellar population in the stellar disk is to study the velocity dispersions \citep{2004A&A...418..989N, 2021MNRAS.506.1761S}. The idea is that stars that formed in the stellar disk and via encounters with other objects, their velocities change over time such that older stars have more “heated” orbits. Using the largest kinematic-only (i.e. no elemental abundances) sample of RR~Lyrae variables it has been shown that their velocity dispersions are consistent with models indicating that the sample has an age of about 2 billion years \citep{2021MNRAS.502.5686I}. This finding appears incompatible with our new accreted disk stellar population. However, what is measured based on the kinematics is not the age of the stars but the age of the population since it became part of a cool stellar disk. Thus, we argue that the two findings are compatible in a scenario where a stellar population consisting of stars formed up to 10 billion years ago merged with the Milky Way stellar disk about 2 billion years ago. 
The population was dragged into the plane, onto very circular orbits and has since been heated. 

Is such a scenario plausible? Simulations show that it is possible to accrete stellar systems into the Milky Way and, given the right inclination of the merger, via plane-dragging deposit the stars into the stellar disk on disk-like, essentially circular orbits \citep{2008MNRAS.389.1041R, 2017MNRAS.472.3722G}. However, it is difficult to make a quantitative comparison between our data and such models due to the small observational sample and the resolution of the models.

\citet{2020ApJ...902..119D} present evidence from stellar overdensity shell structures of a dwarf galaxy passage through the Milky Way center 2.7 billion years ago, suggesting the progenitor galaxy could have been the {\it Gaia}-Sausage-Enceladus or another unknown galaxy. { Such timing is compatible with the "heating age" of the RR~Lyrae.}

It is difficult to assess the size of the galaxy that merged with ours and later deposited its stars into the local stellar disk. Mass-metallicity relations that evolve with redshift would, because of the high metallicity of the stars, indicate a merger as massive as M$_{\star} = 10^{11}$ to $10^{12}$  M$_{\odot}$ \citep{2016MNRAS.456.2140M}. This assumes that the merging galaxy retained all of its stars at the time of the merger. Assuming that the merger took place about 2 billion years ago and the stars are up to 10 billion years old, this leaves about 8 billion years of evolution from formation to merger. This means that the merging galaxy continued to evolve in the Local Group and thanks to its proximity to the Milky Way it may have been stripped of some of its stars and only a smaller portion of it merged with the Milky Way disk.

\section{Conclusions and outlook}

Using elemental abundances from APOGEE DR17 \citep{2016AN....337..863M, 2022ApJS..259...35A} we have identified sample of stars that have chemical signatures of an accreted origin and also have kinematics similar to the Milky Way stellar disk. We have further identified a small sample of RR~Lyrae variables that share the properties with this possibly-accreted component. The RR~Lyrae variables offer the possibility to obtain an independent age estimate of this new stellar component. We find that most likely we are looking at an ancient stellar population that formed outside the Milky Way and was later accreted and dragged into the stellar disk. Only by combining elemental abundances, kinematics and different age dating techniques are we able to confirm its existence.

Looking forward there are both on-going and up-coming large massive spectroscopic surveys that will enable us to identify larger samples of stars with chemical signatures of an accreted origin. In particular, the 4MOST survey will target several million disk stars, including a specific survey to target RR~Lyrae stars. At the same time, theoretical models of aluminum enrichment are needed to improve our understanding of how a galactic system might maintain low [Al/Fe] at high [Fe/H].

\section*{Acknowledgements}
We thank the anonymous referee for their helpful comments and suggestions. 
We thank Eric Andersson, Oscar Agertz, and Justin Read for their contributions to this work and discussions about galaxy mergers.
Henrik J\"onsson is thanked for his guidance on inspection of the APOGEE spectra and unknown lines.

DF, SF, and CLS were partly supported by the grant 2016-03412 from the Swedish Research Council.
TB was supported by grant 2018-04857 from the Swedish Research Council.

This research has made use of NASA's Astrophysics Data System, the SIMBAD database and the VizieR catalogue access tool, operated at CDS, Strasbourg, France, matplotlib, a Python library for publication quality graphics \citep{Hunter:2007}, TOPCAT, an interactive graphical viewer and editor for tabular data \citep{2005ASPC..347...29T}, Astropy, a community-developed core Python package for Astronomy \citep{2018AJ....156..123A, 2013A&A...558A..33A}, pandas \citep{McKinney_2010, McKinney_2011}, and NumPy \citep{harris2020array}. This work has made use of the data from the European Space Agency (ESA) mission {\it Gaia} (\url{https://www.cosmos.esa.int/gaia}), processed by the {\it Gaia} Data Processing and Analysis Consortium (DPAC, \url{https://www.cosmos.esa.int/web/gaia/dpac/consortium}). Funding for the DPAC has been provided by national institutions, in particular the institutions participating in the {\it Gaia} Multilateral Agreement. The acknowledgements were compiled using the Astronomy Acknowledgement Generator.

Funding for the Sloan Digital Sky 
Survey IV has been provided by the 
Alfred P. Sloan Foundation, the U.S. 
Department of Energy Office of 
Science, and the Participating 
Institutions. 

SDSS-IV acknowledges support and 
resources from the Center for High 
Performance Computing  at the 
University of Utah. The SDSS 
website is www.sdss.org.

SDSS-IV is managed by the 
Astrophysical Research Consortium 
for the Participating Institutions 
of the SDSS Collaboration including 
the Brazilian Participation Group, 
the Carnegie Institution for Science, 
Carnegie Mellon University, Center for 
Astrophysics | Harvard \& 
Smithsonian, the Chilean Participation 
Group, the French Participation Group, 
Instituto de Astrof\'isica de 
Canarias, The Johns Hopkins 
University, Kavli Institute for the 
Physics and Mathematics of the 
Universe (IPMU) / University of 
Tokyo, the Korean Participation Group, 
Lawrence Berkeley National Laboratory, 
Leibniz Institut f\"ur Astrophysik 
Potsdam (AIP),  Max-Planck-Institut 
f\"ur Astronomie (MPIA Heidelberg), 
Max-Planck-Institut f\"ur 
Astrophysik (MPA Garching), 
Max-Planck-Institut f\"ur 
Extraterrestrische Physik (MPE), 
National Astronomical Observatories of 
China, New Mexico State University, 
New York University, University of 
Notre Dame, Observat\'ario 
Nacional / MCTI, The Ohio State 
University, Pennsylvania State 
University, Shanghai 
Astronomical Observatory, United 
Kingdom Participation Group, 
Universidad Nacional Aut\'onoma 
de M\'exico, University of Arizona, 
University of Colorado Boulder, 
University of Oxford, University of 
Portsmouth, University of Utah, 
University of Virginia, University 
of Washington, University of 
Wisconsin, Vanderbilt University, 
and Yale University.


\bibliographystyle{aasjournal}
\bibliography{references}

\label{lastpage}
\end{document}